\documentclass[prl, reprint, superscriptaddress]{revtex4-1}
\usepackage{amsmath}
\usepackage{amssymb}
\usepackage{bm}
\usepackage{graphicx}
\usepackage{microtype}
\usepackage{units} 

\renewcommand{\vec}[1]{\bm{#1}}

\newcommand{\avr}[1]{\left\langle#1\right\rangle}

\DeclareMathOperator{\Pe}{Pe}

\begin{document}
\title{Rheology of inelastic hard spheres at finite density and shear rate}
\author{W. Till Kranz} 
\email{kranz@thp.uni-koeln.de}
\affiliation{Institut f\"ur Materialphysik im Weltraum, Deutsches
  Zentrum f\"ur Luft- und Raumfahrt, 51170 K\"oln, Germany}
\affiliation{Institut f\"ur Theoretische Physik, Universit\"at zu
  K\"oln, 50937 K\"oln, Germany}
\author{Fabian Frahsa}
\affiliation{Fachbereich Physik, Universit\"at Konstanz, 78457
  Konstanz, Germany}
\author{Annette Zippelius}
\affiliation{Institut f\"ur Theoretische Physik,
  Georg-August-Universit\"at G\"ottingen, 37077 G\"ottingen, Germany}
\author{Matthias Fuchs}
\affiliation{Fachbereich Physik, Universit\"at Konstanz, 78457
  Konstanz, Germany}
\author{Matthias Sperl}
\affiliation{Institut f\"ur Materialphysik im Weltraum, Deutsches
  Zentrum f\"ur Luft- und Raumfahrt, 51170 K\"oln, Germany}
\affiliation{Institut f\"ur Theoretische Physik, Universit\"at zu
  K\"oln, 50937 K\"oln, Germany}

\begin{abstract}
  Considering a granular fluid of inelastic smooth hard spheres we discuss the
  conditions delineating the rheological regimes comprising Newtonian,
  Bagnoldian, shear thinning, and shear thickening behavior. Developing a
  kinetic theory, valid at finite shear rates and densities around the glass
  transition density, we predict the viscosity and Bagnold coefficient at
  practically relevant values of the control parameters. The determination of
  full flow curves relating the shear stress $\sigma$ to the shear rate
  $\dot\gamma$, and predictions of the yield stress complete our discussion of
  granular rheology derived from first principles.
\end{abstract}


\maketitle

Predicting and understanding granular flow is desirable for safety and
efficiency. Many geophysical flows from avalanches to landslides
involve macroscopic particles and potentially threaten lives all
around the planet \cite{vanwesten+vanasch06,holler07}. A large
fraction of the raw materials handled in industry comes in granular
form \cite{richard+nicodemi05}. With the advent of 3d printing
technologies, this fraction will further increase
\cite{gebhardt12,fateri+gebhardt14}. Hence, efficient handling of
granular flows promises considerable advantages like energy savings
\cite{cleary+sawley02}. The crucial property of granular
particles---namely that they are of macroscopic size---makes a
theoretical description challenging
\cite{jaeger+nagel92,kadanoff99,rajchenbach00}. Firstly, dissipative
collisions break time-reversal symmetry and place granular flows
firmly in the realm of far-from-equilibrium physics
\cite{ausloos+lambiotte05,marconi+puglisi08}. Secondly, the
macroscopic mass of an individual granular particle makes its thermal
excitations negligible \cite{jaeger+nagel92} and necessitates a
driving force that is continuously acting on the granular assembly to
keep it flowing \cite{prevost+egolf02,ojha+lemieux04,abate+durian06}.

At small volume fractions $\varphi\ll1$ and infinitesimal shear rates
$\dot\gamma\to0$ standard procedures starting from the Boltzmann or Enskog
equation predict a Newtonian rheology for (smooth) granular particles. Various
approximate expressions exist for the viscosity as a function of inelasticity
(often quantified by a coefficient of restitution $\varepsilon$) and volume
fraction
\cite{savage+jeffrey81,jenkins+savage83,brey+dufty98,sela+goldhirsch98,garzo+montanero02,brilliantov+poeschel10}.
However, most natural and industrial granular flows occur at considerable
volume fractions all the way up to close packing densities. Relevant shear
rates $\dot\gamma$ are also often substantial compared to microscopic time
scales.

\begin{figure}
  \centering
  \includegraphics[width=\linewidth]{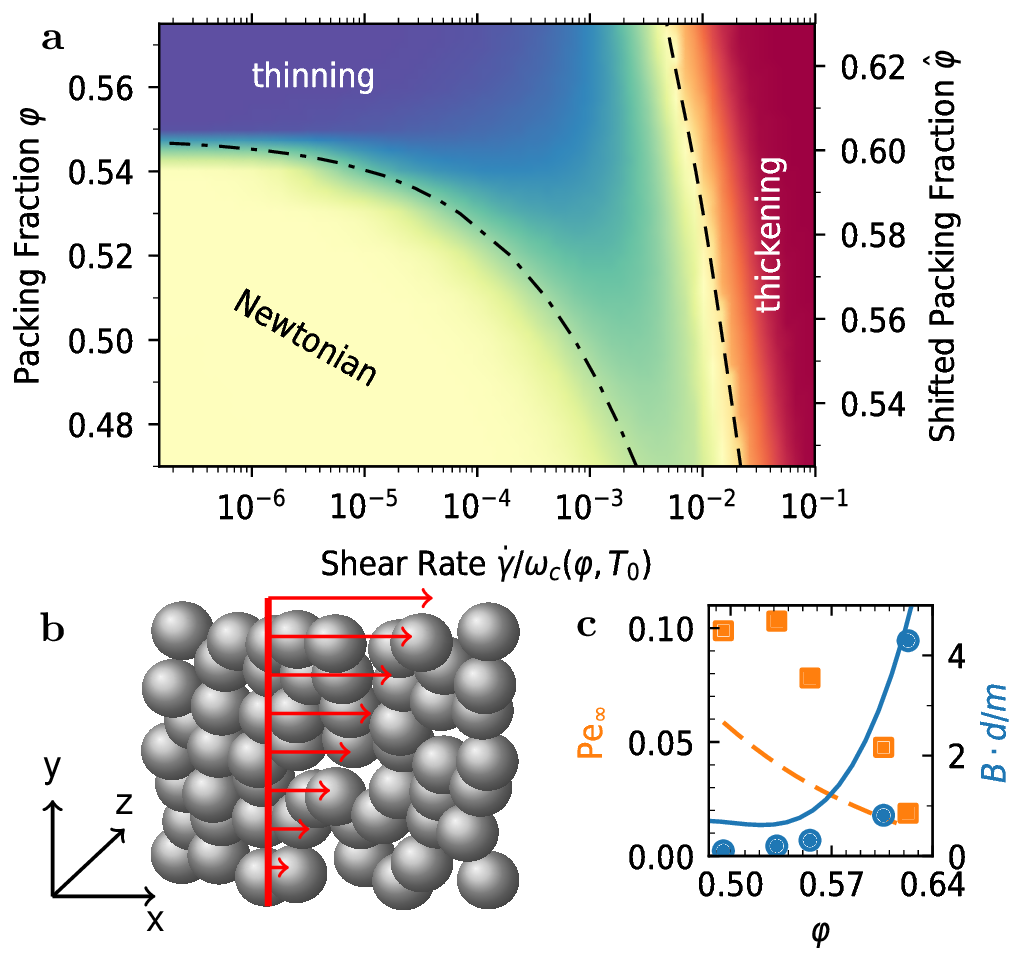}
  \caption{(a) Theoretical dynamic state diagram of a granular fluid with
    coefficient of restitution $\varepsilon=0.5$ depending on packing fraction
    $\varphi$ ($\hat\varphi := \varphi + 5.5\%$) and shear rate
    $\dot\gamma$. Rheological behavior (R) color coded as indicated. Critical
    P\'eclet numbers $\Pe^*$ (dashed) where shear heating becomes important and
    $\Pe_{\alpha}$ (dash-dotted), where $\dot\gamma$, matches the intrinsic
    relaxation rate, $\tau_{\alpha}^{-1}$. (b) Schematic of the shear geometry
    with the shear profile overlayed (red). (c) Maximal P\'eclet number,
    $\Pe_{\infty}$, (orange dashed, left axis), and Bagnold coefficient, $B$,
    (blue solid, right axis) as function of volume fraction $\varphi$
    comparing Bagnold's measurements \cite{bagnold54} (symbols) and the
    kinetic theory presented here (lines).}
  \label{fig:regimes}
\end{figure}

Experimental as well as numerical studies have found a wealth of
phenomena at finite densities and shear rates
\cite{bagnold54,campbell06,hinrichsen+wolf04,lemaitre+roux09,kawasaki+ikeda14,brown+jaeger14}. One
of the earliest results of granular physics by Bagnold
\cite{bagnold54}---now commonly referred to as \emph{Bagnold
  scaling}---is the observation that granular fluids do not follow a
Newtonian rheology, but that the shear stress shows a quadratic
dependence on the shear rate, $\sigma = B{\dot\gamma}^2$,
instead. Phenomenologically, Bagnold scaling is a manifestation of
\emph{shear thickening} as the shear rate dependent viscosity
$\eta(\dot\gamma) \equiv \sigma/\dot\gamma \sim \dot\gamma$ increases
with shear rate. Theoretical predictions of the corresponding Bagnold
coefficient $B$ are rare
\cite{savage+jeffrey81,mitarai+nakanishi05,reddy+kumaran07} and for
low density or low shear rate only. Later studies also show the
opposite behavior, i.e., \emph{shear thinning}, in granular fluids.
Taken together, these observations imply that the Newtonian rheology
is valid only in a limited part of the parameter space and capturing
the full rheology in a single theoretical framework remains a
challenge.

In this letter we will show that, indeed, granular flows display all three
regimes: Newtonian, shear thinning, and shear thickening
(Fig.~\ref{fig:regimes}a). We will derive the conditions to observe any of
these behaviors. Based on this classification, we then calculate the relevant
material properties, namely the Newtonian viscosity $\eta$, the Bagnold
coefficient $B$, and more generally, flow curves, \textit{i.e.},
$\sigma(\dot\gamma)$, or, equivalently, $\eta(\dot\gamma)$.

Formally, the stationary shear stress $\sigma$ at a constant finite shear rate
can be expressed via a \emph{Generalized Green-Kubo Relation},
\begin{equation}
  \label{eq:1}
  \sigma = \frac{\dot\gamma}{VT}\int_0^{\infty}\mathrm{d}t\avr{\sigma_{xy}|\sigma_{xy}(t)}_{\mathrm{ref}},
\end{equation}
as the time integral over the shear stress auto-correlation whose
evolution depends on the flow \cite{fuchs+cates02}. Here $\sigma_{xy}$
is the Kirkwood shear stress defined in terms of the particle
positions and momenta \cite{dufty12}, $V$ is the volume and $T$ the
temperature of a reference state. The average
$\avr\cdot_{\mathrm{ref}}$ is performed with respect to the unsheared
reference system. Although the relation was originally derived for a
reference system in thermal equilibrium \cite{fuchs+cates02} it
ultimately relates stationary states including out-of-equilibrium
reference systems \cite{kranz+frahsa17b}. Our choice allows to specify
the perturbing stress in Eq.~(\ref{eq:1}) microscopically.  The
auto-correlation function can be controlled by identifying a dominant
decay channel. For thermal colloidal suspensions the slow structural
relaxations close to the glass transition provides such a clearly
defined, slow decay \cite{brader+voigtmann09}. Approximating the
stress auto-correlation function in Eq.~(\ref{eq:1}) in terms of the
density correlator $\Phi_{\vec q}(t)$ lies at the heart of the
\emph{Integration Through Transients} (\textsc{ITT}) formalism
\cite{fuchs+cates02,fuchs+cates09},
\begin{equation}
  \label{eq:2}
  \sigma \simeq \dot\gamma\sum_{\vec q}\int_0^{\infty}\mathrm{d}t\mathcal
  V_{\vec q\vec q(-t)}\Phi_{\vec q(-t)}^2(t).
\end{equation}
Here $\vec q(-t) \equiv (\mathsf{1} + \mathsf{k}t)\cdot\vec q$ denotes
the advected wave vector and the velocity gradient tensor
$\mathsf{k}_{\alpha\beta} := \dot\gamma\delta_{\alpha y}\delta_{\beta
  x}$
prescribes simple shear (Fig.~\ref{fig:regimes}b). Note that the
advected wave vector's dependence on $\dot\gamma$ effects a nonlinear
stress-shear rate relation in Eq.~(\ref{eq:2}). The coupling constant
$\mathcal V_{\vec q\vec q(-t)}$ can be calculated explicitly [see
Refs.~\cite{fuchs+cates09,kranz+frahsa17b} and Eq.~(\ref{eq:5})
below]. The \textsc{ITT} framework has led to a wealth of qualitative
and quantitative predictions regarding the rheology of thermalized
colloidal suspensions
\cite{brader+voigtmann07,brader10,amann+siebenbuerger13,nicolas+fuchs16}

\begin{table}[t]
  \caption{Definition of the critical P\'eclet numbers delineating the
    rheological regimes as functions of packing fraction $\varphi$ and
    coefficient of restitution $\varepsilon$. See text for details.}
  \begin{ruledtabular}
    \begin{tabular}{l}
      $\Pe_{\alpha}(\varphi, \epsilon) = (\omega_c\tau_{\alpha})^{-1}$\\
      $\Pe^*(\varphi, \epsilon) = \sqrt2\Gamma(\varepsilon)/2\hat\sigma(\Pe^*, \varphi, \varepsilon)$\\
      $\Pe_{\infty}(\varphi, \epsilon) = \Gamma(\varepsilon)/\hat\sigma(\Pe_{\infty}, \varphi, \varepsilon)$    
    \end{tabular}
  \end{ruledtabular}
  \label{tab:peclet}
\end{table}

\emph{Power Balance}---The temperature $T$ of an overdamped colloidal
suspension is assumed to be controlled by a heat bath. In particular,
one assumes that the work performed on the suspension by the shear
force does \emph{not} increase the temperature but that instead the
temperature can be chosen freely. Formulating \textsc{ITT} for
underdamped, Newtonian dynamics, a thermostat has to be included
explicitly with the final results depending on the precise choice of
artificial thermostat \cite{chong+kim09,suzuki+hayakawa13}. In
granular flows the balance between shear heating and dissipation
occurs naturally and actually controls the qualitative behavior of the
granular fluid. The system we have in mind in the following is a
sheared fluidized bed \cite{bocquet+losert01}. To be specific, let's
consider a fluid composed of monodisperse smooth hard spheres of
diameter $d$, and mass $m=1$ with a coefficient of normal restitution
$\varepsilon$. We assume that, initially, the fluid is prepared at a
given density $n$, or packing fraction $\varphi = \pi nd^3/6$ and a
random fluidization force is applied throughout the system with a
characteristic power per particle $P_D$ to mimic fluidization
\cite{bizon+shattuck99} \footnote{In practice,
  $P_D = P_D(X, \varphi, \dot\gamma)$ is not only a function of the
  control parameters of the driving, $X$, (\textit{e.g.} flow rate,
  shaking strength, etc.) but also depends on the state of the system
  ($\varphi, \dot\gamma$). Here we assume $P_D$ fixed for
  simplicity.}. Then the initial granular temperature $T_0$ results
from the power balance $P_D = \Gamma\omega_c(T_0)T_0$ where
$\Gamma := (1 - \varepsilon^2)/3$.  The collision frequency,
$\omega_c(\varphi, T) \propto \varphi\chi(\varphi)\sqrt T$, increases
with density and temperature and $\chi(\varphi)$ denotes the pair
correlation function at contact \cite{brilliantov+poeschel10}.

Once shear is applied with a prescribed shear rate $\dot\gamma$, shear heating
has to be included in the power balance of the steady state,
\begin{equation}
  \label{eq:3}
  \sigma\dot\gamma + nP_D = n\Gamma\omega_c(T)T,
\end{equation}
resulting in a higher temperature $T > T_0$ (\emph{Protocol H} in
Ref.~\cite{kranz+frahsa17b}). This clearly defines two regimes: A
\emph{fluidization dominated} regime, where
$\sigma\dot\gamma \ll nP_D$ and a \emph{shear dominated} regime, where
$\sigma\dot\gamma \gg nP_D$ including purely shear driven systems
($P_D\equiv0$).  Choosing the particle diameter, $d$, as our length
scale and the inverse collision frequency in the stationary state,
$\omega_c^{-1}(T)$, as our time scale the packing fraction, the
coefficient of restitution, and the P\'eclet number,
$\Pe = \dot\gamma/\omega_c(T)$, alone determine the system's
state. The temperature only controls the overall timescale.  The shear
stress scales as $\sigma = nT\hat\sigma$ with a dimensionless function
$\hat\sigma = \hat\sigma(\Pe, \varphi, \epsilon)$ because $nT$
provides the only scale for an energy density. The crossover between
the regimes is expected where $\sigma\dot\gamma\sim nP_D$. This
implicitly determines a critical P\'eclet number
$\Pe^*(\varphi, \varepsilon)$ (cf.\ Tab.~\ref{tab:peclet}).

\begin{figure*}[t]
  \includegraphics{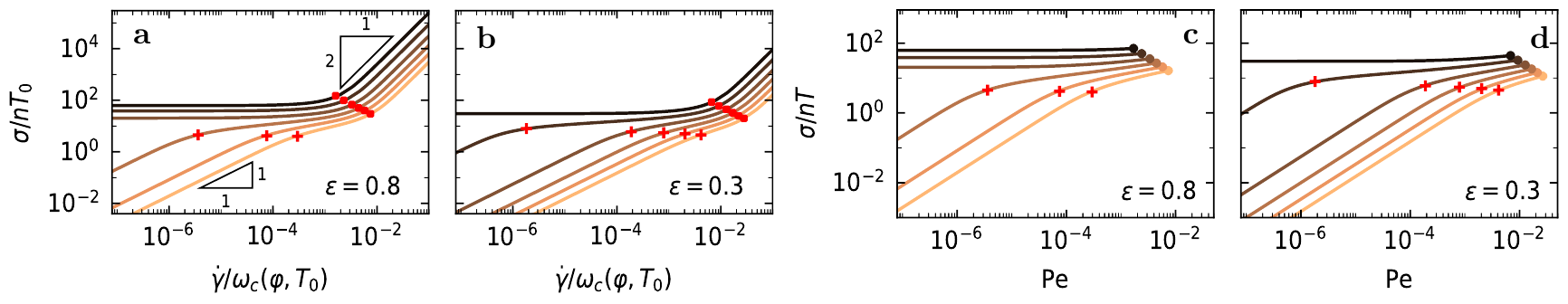}
  \caption{Flow curves: Shear stress $\sigma$ as a function of shear
    rate $\dot\gamma$ (a, b) or P\'eclet number $\Pe$ (c,d) for a number
    of packing fractions and coefficient of restitution
    $\varepsilon=0.8$ (a,c), and $\varepsilon = 0.3$ (b,d). The
    packing fraction increases from bottom ($\varphi=0.48$) to top
    ($\varphi=0.58$). Critical P\'eclet numbers indicated by symbols:
    $\Pe_{\alpha}$ (+), $\Pe^*$ [$\blacksquare$, omitted in (c,d) for
    clarity], and $\Pe_{\infty}$ (\textbullet).}
  \label{fig:flow}
\end{figure*}

In the fluidization dominated regime, the temperature remains
independent of the shear rate such that we can linearize
$\hat\sigma \sim \dot\gamma$ to find Newtonian behavior. In the shear
dominated regime the power balance reads
$nT\hat\sigma\dot\gamma = nT\Gamma\omega_c$ \cite{lemaitre02},
\textit{i.e.}, the collision frequency is proportional to the shear
rate and the corresponding P\'eclet number,
$\Pe = \Pe_{\infty}(\varphi, \varepsilon) > 0$ (cf.\
Tab.~\ref{tab:peclet}), is independent of the shear rate. From
$\dot\gamma\propto\omega_c\propto\sqrt T$ we obtain
$T \propto {\dot\gamma}^2$ and with that,
$\sigma \propto T \propto {\dot\gamma}^2$: Bagnold rheology is
observed in the shear dominated regime where the shear rate is the
only time scale and the P\'eclet number, $\Pe_{\infty}$, is fixed by
the packing fraction and the coefficient of restitution. Upon
increasing the shear rate $\dot\gamma$ towards the Bagnold regime, the
temperature $T(\Pe) = T_0\cdot(1 - \Pe/\Pe_{\infty})^{-2/3}$
diverges. This implies that $\Pe_{\infty}$ is the maximal P\'eclet
number which can not be exceeded in a granular fluid.

\emph{Glassy dynamics and yield stress}---Granular fluids have been
found to undergo a glass transition \cite{abate+durian06} at a
critical packing fraction $\varphi_c(\varepsilon)$ which increases
with increasing dissipation \cite{kranz+sperl10}. Upon approaching the
(granular) glass transition, $\varphi\nearrow\varphi_c$, the
characteristic correlation time for density fluctuations,
$\tau_{\alpha}$, diverges. For $\dot\gamma \ll \tau_{\alpha}^{-1}$ the
rheology is Newtonian, as the granular fluid can respond immediately
to the slow shear deformations. However, the viscosity diverges as
$\varphi\nearrow\varphi_c$. For higher prescribed shear rates,
$\dot\gamma \gtrsim \tau_{\alpha}^{-1}$ [\textit{i.e.},
$\Pe\gtrsim\Pe_{\alpha}(\varphi, \varepsilon)$, cf.\
Tab.~\ref{tab:peclet}], the glass is forcibly molten.

To lowest order it can be assumed that $\Phi_{\vec q}(t\to\infty) \propto
e^{-\dot\gamma t}$ for $\Pe > \Pe_{\alpha}$. Consequently the
Green-Kubo-Integral, Eq.~(\ref{eq:2}), becomes independent of the shear rate,
\textit{i.e}, $\sigma(\dot\gamma) \approx$ const. In terms of the viscosity
$\eta(\dot\gamma) \sim 1/\dot\gamma$ we expect to observe shear thinning. For
even higher shear rates, $\Pe > \Pe^*$, shear heating will dominate and
eventually bend the flow curve to the shear thickening Bagnold regime.  For
densities above the glass transition, $\varphi > \varphi_c$, $\Pe_{\alpha}
\to0$ and the Newtonian regime vanishes altogether. Instead, a finite
(dynamical) yield stress $\sigma_y := \sigma(\dot\gamma\to0) > 0$ emerges
which has to be overcome to melt the amorphous granular glass.

\textit{Granular Integration Through Transients}---Recently
\cite{kranz+frahsa17b,kranz+sperl17}, we derived a non-equilibrium
\textsc{ITT} formalism for granular fluids (g\textsc{ITT}) employing
granular mode-coupling theory (\textsc{MCT})
\cite{kranz+sperl10,kranz+sperl13}. As a central result, we obtain an
expression for the generalized Green-Kubo relation in isotropic
approximation,
\begin{multline}
  \label{eq:5}
  \frac{\sigma}{nT} = \frac{\sigma_0}{nT} 
  + \Pe\frac{1 + \varepsilon}{2\varphi}
  \int_0^{\infty}\frac{\mathrm{d}\tau}{\sqrt{1 + (\Pe\tau)^2/3}}\\
  \times\int_0^{\infty}\frac{\mathrm{d}q^*q^{*4}}{360\pi}
  \times\frac{S'_{q^*(-\tau)}S'_{q^*}}{S_{q^*}^2}
  \Phi_{q^*(-\tau)}^2(\tau),
\end{multline}
extending the low-density, Enskog prediction $\sigma_0$
\cite{garzo+montanero02}. Here $S'_q$ denotes the derivative of the static
structure factor, $q^* := qd$, and $\tau := \omega_ct$, are the dimensionless
wave number and time, respectively, and $q^*(-\tau) = q^*\sqrt{1 +
  (\Pe\tau)^2/3}$

\newpage

\begin{figure*}[t]
  \includegraphics{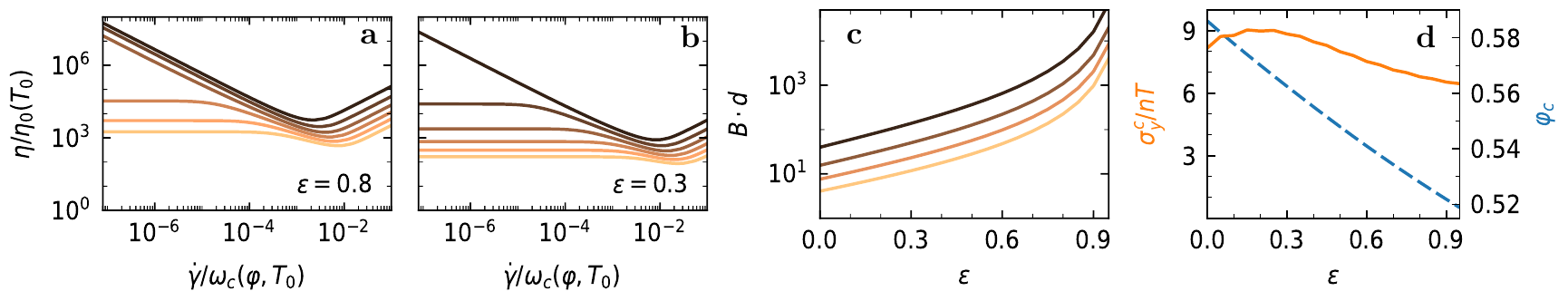}
  \caption{(a, b) Viscosity $\eta(\dot\gamma)$ normalized by the Boltzmann
    viscosity $\eta_0$ as a function of shear rate $\dot\gamma$. Same
    parameters as in Fig.~\ref{fig:flow}. (c) Bagnold coefficient $B$ as a
    function of coefficient of restitution $\epsilon$ for a few packing
    fractions from $\varphi=0.48$ (bottom) to $\varphi=0.57$ (top). (d) Yield
    stress $\sigma_y^c$ (left axis, solid) at the glass transition packing
    fraction $\varphi_c$ (right axis, dashed, from \cite{kranz+sperl10}) as a
    function of the coefficient of restitution $\varepsilon$.}
  \label{fig:eta}
\end{figure*}

Numerically solving Eq. (\ref{eq:5}) together with the granular \textsc{MCT}
equations yields flow curves like shown in Fig.~\ref{fig:flow} as well as the
critical P\'eclet numbers $\Pe_{\alpha}, \Pe^*$, and $\Pe_{\infty}$
(Fig.~\ref{fig:regimes}a). For more details, see
Ref.~\cite{kranz+frahsa17b}. Indeed, we observe all the regimes discussed
above: (i) Newtonian behavior, $\sigma = \eta\dot\gamma$, for low densities
and shear rates, $\Pe < \Pe_{\alpha}$, (ii) a yield stress above the glass
transition density, and generally thinning for $\Pe_{\alpha} < \Pe < \Pe^*$,
and (iii) the Bagnold regime, $\sigma = B{\dot\gamma}^2$, for large shear
rates, $\Pe > \Pe^*$ where the flow curves end at $\Pe = \Pe_{\infty}$
[Fig.~\ref{fig:flow}(c,d)].  Considering the generalized viscosity
$\eta(\dot\gamma)$ [Fig.~\ref{fig:eta}(a,b)] makes the thinning and thickening
terminology particularly transparent. Note that we span many orders of
magnitude in shear rate and viscosity. The critical density's dependence on
$\varepsilon$ (Fig.~\ref{fig:eta}d), $\varphi_c(\epsilon)$, strongly
influences the flow behavior at fixed flow conditions if the coefficient of
restitution is varied (cf.\ Figs.~\ref{fig:flow}, \ref{fig:eta}). The
rheological regimes can be classified by $R
:= \partial\ln\eta(\dot\gamma)/\partial\ln\dot\gamma$ (color coded in
Fig.~\ref{fig:regimes}a). Here, $R = 0$ corresponds to Newtonian rheology and
$R > 0$ ($R < 0$) to shear thickening (thinning) behavior
\footnote{\textsc{MCT} using the standard approximations is known to
  underestimate the glass transition density for elastic hard spheres
  ($\varepsilon=1$), $\varphi_c^{\mathrm{MCT}} \approx 0.516$
  \cite{bengtzelius+goetze84}, by about 5.5\% compared to the experimental
  value $\varphi_c^{\mathrm{exp}} \approx 0.57$
  \cite{vanmegen+mortensen98}. Assuming that this difference remains constant
  for $\varepsilon < 1$, we add a tentatively shifted axis $\hat\varphi =
  \varphi + 5.5\%$ to Fig.~\ref{fig:regimes}a to facilitate comparison to
  experiments.}.

\textit{Slow shear}---For small shear rates $\Pe\to0$ in the linear response
regime, the glass transition divides the state space into two qualitatively
different phases. For relatively low densities, $\varphi < \varphi_c$,
g\textsc{ITT} provides corrections of order $\varphi^2$ to the low density,
Enskog predictions \cite{garzo+montanero02} for the viscosity. For densities
approaching the glass transition, the divergence of the relaxation time,
$\tau_{\alpha}$, results in a divergent viscosity,
$\eta(\varphi\nearrow\varphi_c) \propto [\varphi_c(\varepsilon) -
\varphi]^{-\gamma(\varepsilon)}$.  The critical exponent $\gamma(\varepsilon)
\sim 2.5$--$2.3$ weakly decreases with increasing dissipation
\cite{kranz+sperl13} and compares well with the experimental value $\gamma
\approx 2.35$ \cite{bonnoit+darnige10}. Experimental values for
$\varphi_c(\varepsilon)$
\cite{ovarlez+bertrand06,brown+jaeger09,gamonpilas+morris16} are compatible
with our interpretation as a granular glass transition as well. At the glass
transition, $\varphi\equiv\varphi_c(\varepsilon)$, we find a finite critical
yield stress (cf.\ Fig.~\ref{fig:eta}d), $\sigma_y^c$, in the range
$6nT$--$9nT$, compatible with experiments \cite{brown+jaeger09}, and
comparable to theoretical predictions ($\sigma_y^c/nT\sim 6$
\cite{fuchs+cates03}) and measurements ($\sigma_y^c/nT\sim10$--$15$
\cite{petekidis+vlassopoulos04}) for colloidal suspensions. Thereby placing
granular fluids firmly in the realm of \emph{soft matter}.

\textit{Fast shear}---For high P\'eclet numbers $\Pe \to \Pe_{\infty}$,
Eq.~(\ref{eq:5}) can be used to predict the Bagnold coefficient, $B$ (cf.\
Fig.~\ref{fig:eta}c). The latter increases with density and elasticity (larger
$\varepsilon$) as both trends makes the temperature more sensitive to changes
in the shear rate. In particular, $B$ diverges as $\varepsilon\to1$.  

As a first quantitative application of g\textsc{ITT}, let us compare
our predictions to Bagnold's original data \cite{bagnold54}. To this
end, we extract the Bagnold coefficient, $B$, and the maximal P\'eclet
number, $\Pe_{\infty}$, from his measurements
(Fig.~\ref{fig:regimes}c) \footnote{For details and additional
  analysis of Ref.~\cite{fall+lemaitre10} see Supplemental Material at
  xxx including Refs.~\cite{bannerman+lue10,lorenz+tuozzolo97}}. Our
kinetic theory proves to recover the qualitative trends of both $B$
and $\Pe_{\infty}$ as a function of packing fraction with no
adjustable parameters. Considering that Bagnold's measurements have
been shown to leave room for improvement \cite{hunt+zenit02} the
prediction of g\textsc{ITT} compare favorably.  Qualitatively the
strongly agitated flow curves of
Refs.~\cite{dijksman+wortel11,wortel+dijksman14} agree well with our
discussion showing all three regimes. However, note that the packing
fraction in the shear zone is unknown and uncontrolled in these
experiments in contrast to what is assumed here. More tailored and
careful measurements are needed to assess g\textsc{ITT}'s quantitative
accuracy.

Depending on experimental conditions additional effects may become
relevant that go beyond the model considered in this letter. On earth
(but not, e.g., on the moon \cite{kokelaar+bahia17}) all granular
flows are actually two-phase flows of granular particles together with
an interstitial fluid (mostly air or water). At sufficiently low
packing fraction, the effective viscosity of the molecular fluid
becomes relevant \cite{fall+lemaitre10}. This will introduce another
Newtonian regime at small shear rates that crosses over to Bagnold
rheology when the stress induced by the granular particles dominates
over the viscous stress of the interstitial fluid. For high shear
rates, the Bagnold regime will obviously not extend to
$\dot\gamma\to \infty$. At some point typical inter-particle forces
are so high that interactions can no longer be regarded as
hard-core. A finite interaction time compared to the shear rate
appears as a new time scale and destroys Bagnold scaling
\cite{vagberg+olsson17}.  For very high densities, approaching random
close packing, the rheology will be dominated by the imminent jamming
transition which is not accounted for in the present model.  We
prescribe a linear shear profile and a homogeneous constant density
and therefore necessarily obtain monotonous flow
curves. Non-monotonic, unstable flow curves and the associated
discontinuous shear thickening are possible in inhomogeneous systems
only
\cite{brown+jaeger14,dijksman+wortel11,wortel+dijksman14,gnoli+lasanta16}.

\textit{Conclusion}---To summarize, we have discussed that the rich rheology
of a granular fluid is controlled by three critical P\'eclet numbers
(Tab.~\ref{tab:peclet}): (i) Newtonian rheology prevails for $\Pe <
\Pe_{\alpha}$.  (ii) For intermediate shear rates, $\Pe_{\alpha} < \Pe <
\Pe^*$, shear thinning reflects that the shear rate is faster than the
intrinsic relaxation rate of the fluid which eventually results in a finite
dynamic yield stress above the glass transition density. The latter can be
substantially lower than the jamming transition commonly located at random
close packing $\varphi_{\mathrm{rcp}}\approx0.64$ \cite{ohern+langer2002}. For
low densities $\Pe_{\alpha}\sim1$ and at the same time $\Pe^*\ll1$ in the
elastic limit $\varepsilon\to1$. Under these conditions, $\Pe_{\alpha} >
\Pe^*$ and the thinning regime vanishes altogether. (iii) For $\Pe^* < \Pe <
\Pe_{\infty}$ strong shear heating leads to shear thickening behavior which
ultimately entails Bagnold scaling as the P\'eclet number approaches its
maximum, $\Pe\to\Pe_{\infty}$. This constitutes yet another shear thickening
mechanism different from other mechanisms discussed in the literature, namely
clustering \cite{melrose+ball04,wagner+brady09}, dilation
\cite{brown+jaeger12,fall+bertrand12}, friction
\cite{heussinger13,guy+hermes15}, or steric effects
\cite{picano+breugem13}. The fact that $\Pe_{\infty}$ is finite implies that a
kinetic theory to predict the Bagnold coefficient $B$ must be applicable at
finite shear rates.

To support our arguments and to make them quantitative, we presented a kinetic
theory based on the \textsc{ITT} formalism that recovers the phenomenology,
covering many orders of magnitude in shear rate and shear stress, or
viscosity, respectively. Earlier attempts at formulating \textsc{ITT} for
inelastic \emph{soft} spheres \cite{suzuki+hayakawa14} retain no dissipative
effects on the same level of approximation. For the inelastic \emph{hard}
sphere fluid considered here, besides the implicit dependence of $\Phi_{\vec
  k}(t)$ on the coefficient of restitution $\varepsilon$ \cite{kranz+sperl13},
Eq.~(\ref{eq:5}) also includes an explicit dependence on
$\varepsilon$. Thereby we extended quantitative predictions for the transport
coefficients of a sheared granular fluid beyond the low density and low shear
rate regime amenable to standard kinetic theories.

The results presented here will be useful as constitutive equations for
modeling and simulating large scale granular flows which demand a continuum
description. In addition, the Bagnold coefficient is needed for a recent
kinetic theory \cite{kumaran14}. We also hope that the availability of a
kinetic theory for granular shear flow in a range of practically relevant
parameters will spur quantitative experiments.

\begin{acknowledgments}
  We acknowledge illuminating discussions with Hisao Hayakawa, Koshiro Suzuki,
  Thomas Voigtmann, and Claus Heussinger. We thank Philip Born for critically
  reading the manuscript and the DFG for partial funding through FOR\ 1394 and
  KR\ 4867/2-1.
\end{acknowledgments}

%

\appendix

\section{Extracting $B$ and $\Pe_{\infty}$ from
  Bagnold~1954}
\label{sec:extr-bagn-coeff}

\begin{figure}[tb]
  \centering
  \includegraphics{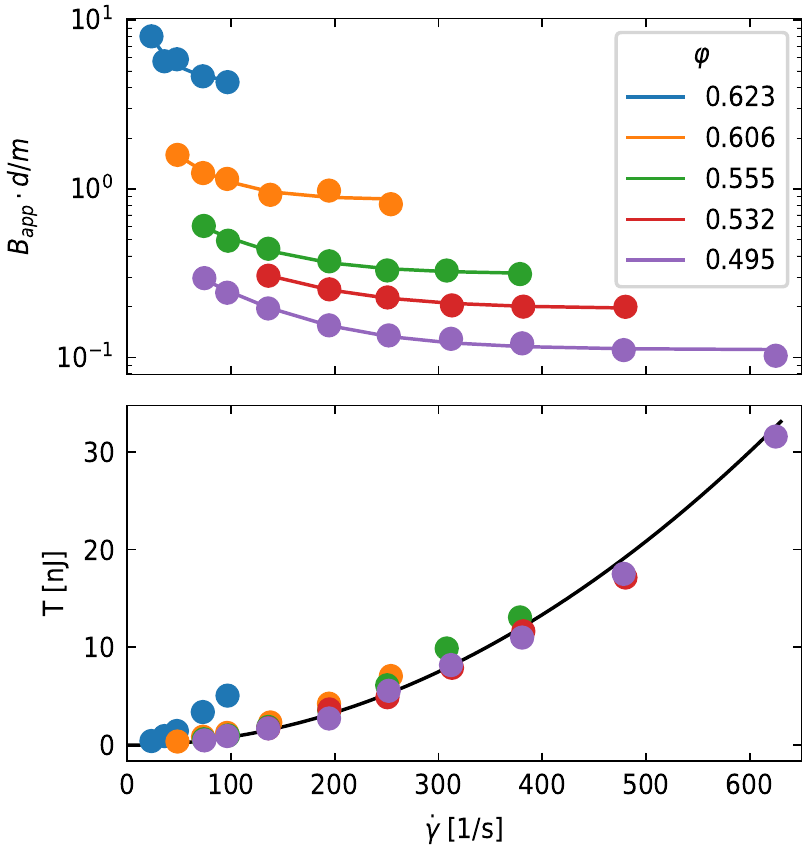}
  \caption{(a) Apparent Bagnold coefficient $B_{\mathrm{app}}$ as a function
    of shear rate $\dot\gamma$ and exponential fits (lines). (b) Granular
    temperature $T$ as a function of shear rate $\dot\gamma$ calculated from
    the pressure data. The black line $\propto{\dot\gamma}^2$ is a guide to
    the eye. Symbols denote data extracted from Ref.~\cite{bagnold54}. Packing
    fractions $\varphi$ as indicated.}
  \label{fig:expfit}
\end{figure}

We read off shear stress and pressure data from Figs.~3 and 4 in
Ref.~\cite{bagnold54}. Calculation of the apparent Bagnold coefficient
is straight forward,
$B_{\mathrm{app}}(\dot\gamma) =
\sigma/{\dot\gamma}^2$.
Fig.~\ref{fig:expfit}a shows that $B_{\mathrm{app}}(\dot\gamma)$
converges exponentially to what we take as the proper Bagnold
coefficient $B \equiv B_{\mathrm{app}}(\dot\gamma\to\infty)$. The
particle mass $m\approx\unit[1.2]{mg}$ we obtain from the particle
diameter $d=\unit[1.32]{mm}$ and the fact that they are density
matched with water. Bagnold reports packing densities in terms of the
linear density $\lambda$ which can directly be converted into a
packing fraction \cite{hunt+zenit02}. Bagnold did not measure the
coefficient of restitution of his particles so we assume
$\varepsilon = 0$ \footnote{Jan Haeberle (private communications)}.

To estimate the P\'eclet number from the shear rate, we need the collision
frequency or the granular temperature. Neither of which have been measured by
Bagnold. We estimate the temperature from the pressure by inverting the
Woodcock equation of state \cite{bannerman+lue10}. The result is shown in
Fig.~\ref{fig:expfit}b. We observe that we reasonably recover Bagnold scaling
$T\propto{\dot\gamma}^2$ and that the highest packing fraction seems to behave
differently from the rest \cite{hunt+zenit02}.

\section{Bagnold coefficient $B$ from Fall et al.}
\label{sec:bagn-coeff-b}

Although the publication has a different focus, Bagnold coefficients may be
extracted from Fig.~2a in Fall \textit{et al.}  \cite{fall+lemaitre10}. From
the given particle diameter $d \approx \unit[40]{\mu m}$ and the polystyrene
particle's mass density $\rho_s = \unit[1.05]{g\cdot cm^3}$, we can derive the
appropriately non-dimensionalized apparent Bagnold coefficient
$B_{\mathrm{app}}$. The coefficient of restitution $\varepsilon$ is not
reported. From the measurements of much larger polystyrene spheres at higher
impact velocities \cite{lorenz+tuozzolo97}, we infer that the smaller
particles at typically slower impact speeds have a coefficient of restitution
$\varepsilon \gtrsim 0.95$.

Applying the same exponential fit as for Bagnold's data, we find that only the
lowest three densities allow to extract a reliable value for the asymptotic
Bagnold coefficient $B$. Treating $\varepsilon$ as a fit parameter, good
agreement between theory and experiment for $\varepsilon = 0.97$ can be
obtained (see Tab.~\ref{tab:lemaitre}). Note that the nearly elastic particles
of Fall \textit{et al.} result in much higher Bagnold coefficients than the
strongly inelastic particles employed by Bagnold as expected.

\begin{table}[t]
  \caption{Dimensionless Bagnold coefficient $B\cdot d/m$ from as measured in
    the experiment of Ref.~\cite{fall+lemaitre10} compared to the g\textsc{ITT} prediction.}
  \centering
  \begin{tabular}{p{3em}|p{4em}c}
    \hline\hline
    $\varphi$ & Exp. & gITT\\\hline
    0.568 & $16\cdot10^3$ & $25\cdot10^3$\\
    0.576 & $25\cdot10^3$ & $32\cdot10^3$\\
    0.581 & $42\cdot10^3$ & $36\cdot10^3$\\
    \hline\hline
  \end{tabular}
  \label{tab:lemaitre}
\end{table}

\end{document}